\documentclass[
    ,final            
  ]
  {aipproc}

\layoutstyle{8x11double}

\begin{document}

\title{Period-Luminosity Relation for Type II Cepheids}

\classification{<Replace this text with PACS numbers; choose from this list:
                \texttt{http://www.aip..org/pacs/index.html}>}

\author{Noriyuki Matsunaga}{
  address={Institute of Astronomy, University of Tokyo; JSPS Research Fellow}
}

\author{Michael W. Feast}{
  address={Department of Astronomy, University of Cape Town}
  ,altaddress={South African Astronomical Observatory}
}

\author{John W. Menzies}{
  address={South African Astronomical Observatory}
}

\begin{abstract}
We have estimated $JHK_{\mathrm{s}}$ magnitudes corrected to mean intensity for
LMC type II Cepheids found in the OGLE-III survey. Period-luminosity 
relations (PLRs) are derived in $JHK_{\mathrm{s}}$ as well as
in a reddening-free $VI$
parameter. The BL Her stars ($P < 4$~d)
and the W Vir stars ($P = 4$ to 20~d) are co-linear in these PLRs.
The slopes of the infrared relations agree with those found previously for
type II Cepheids in globular clusters within the uncertainties.
Using the pulsation parallaxes of V553~Cen and SW~Tau, the data lead to
an LMC modulus of $18.46\pm 0.10$~mag, uncorrected for any metallicity effects.
We have now established the PLR of type II Cepheids as a distance indicator 
by confirming that (almost) the same PLR satisfies the distributions
in the PL diagram of type II Cepheids
in (at least) two different systems, i.e.\  the LMC
and Galactic globular clusters, and by calibrating the zero point of the PLR.
RV Tau stars in the LMC, as a group, are not co-linear
with the shorter-period type II
Cepheids in the infrared PLRs in marked contrast to such stars in
globular clusters. We note differences in period distribution and
infrared colors for RV Tau stars in the LMC, globular clusters  
and Galactic field. We also compare the PLR
of type II Cepheids with that of classical Cepheids.
\end{abstract}

\maketitle


\section{Introduction}

Type II Cepheids (CephIIs) have periods in the same range as classical
Cepheids but are lower-mass stars belonging to disk and halo populations.
In 2006, we showed that CephIIs in globular clusters defined
narrow period-luminosity relations (PLRs) in the near-infrared bands,
$JHK_{\mathrm{s}}$ \cite{Matsunaga-2006}.
This suggests that these stars may be useful distance
indicators for pop.\  II stellar systems. Pulsation parallaxes of
Galactic CephIIs
were used by Feast and collaborators to calibrate these cluster PLRs and
to discuss the distances of the LMC and the Galactic Center \cite{Feast-2008}.
Recently, the OGLE-III survey team catalogued 197 CephIIs in the LMC,
confirming their clear PLR in visible \cite{Soszynski-2008}.
In this contribution, we discuss 
near-IR observations
of these CephIIs and their PLR. We also make a direct comparison
between this PLR and that of classical Cepheids. 

\section{Sample and data analysis}

We searched for near-IR counterparts of the LMC CephIIs \cite{Soszynski-2008}
in the IRSF catalogue
for the LMC \cite{Kato-2007}. This is a point-source catalogue 
in the $JHK_{\mathrm{s}}$ bands obtained
with the 1.4-m IRSF telescope and SIRIUS camera based at SAAO, Sutherland,
South Africa. We found matches for 188 CephIIs with a tolerance of
0.5$^{\prime\prime}$. 
Based on notes given by \cite{Soszynski-2008}, we omitted several stars,
e.g.\  those blended with other stars and those showing eclipses,
from the following discussion regarding the PLR.
Since the OGLE-III catalogue \cite{Soszynski-2008} gives periods and
dates of maximum in $I$, we can derive the phases of the single
$JHK_{\mathrm{s}}$ observations. Assuming that the light curves in the near-IR
bands are similar to those in $I$, we obtained an estimate of the mean
(phase-corrected) magnitude in each IR band.
We confirmed the validity of this method by using the data for CephIIs
which have
two IRSF measurements observed within two neighboring frames
(see \cite{Matsunaga-2009} for details).

\section{Period-Luminosity relation}
\begin{figure}
  \includegraphics[width=.45\textwidth]{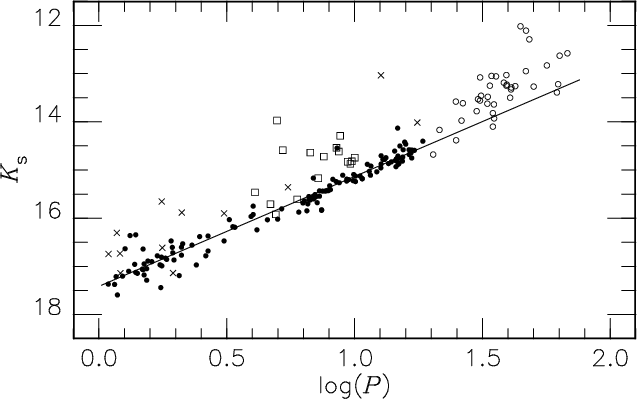}
  \caption{Period-magnitude relation of CephIIs in the LMC.
The phase-corrected $K_{\mathrm{s}}$ magnitudes are plotted against periods.
Filled circles indicate BL Her and W Vir stars used to solve
the PLRs while crosses indicate those excluded,
open squares peculiar W Vir stars,
and open circles RV Tau stars.\label{fig:PLR}}
\end{figure}

Fig.~\ref{fig:PLR} plots the phase-corrected $K_{\mathrm{s}}$
magnitudes against
$\log P$ for the LMC CephIIs. It is evident that the RV Tau stars do not
continue the linear PLRs to longer periods, so that
they are not included in fitting the PLR.
The slopes of the PLR do not differ
significantly from those found by \cite{Matsunaga-2006} for CephIIs
in globular clusters. The scatters around the linear solutions
are compatible with those of the PLRs for the cluster variables.

There are two Galactic BL Her stars with known pulsation parallaxes:
V553~Cen ($\log P =0.314$) and SW~Tau ($\log P = 0.200$) \cite{Feast-2008}.
Using the data of tables 4 and 5 of \cite{Feast-2008},
we can calibrate the zero-points
of the PLRs and hence estimate the modulus of the LMC,
$18.46\pm 0.1$~mag without any metallicity correction.
This agrees
well with those derived from classical Cepheids with trigonometrical
parallaxes by \cite{vanLeeuwen-2007} and \cite{Benedict-2007}.
These authors found $18.52\pm 0.03$ from a PL($W$) relation
and $18.47\pm 0.03$ from a PL($K_{\mathrm{s}}$) relation again without
metallicity corrections.

In the course of our work
we found an important implication for
$\kappa$~Pav.
This star ($\log P = 0.959$) has long been thought of as probably the
nearest CephII and hence a prime candidate for fixing the distance scale
for these objects. 
However, 
its pulsation parallax \cite{Feast-2008}
leads to a distance modulus of 
$6.55 \pm 0.07$~mag  placing it well above the PLRs
in the optical and near-IR
and in the region of
the peculiar W Vir stars which was identified by S08.
Further evidence that $\kappa$~Pav belongs to the peculiar W Vir
class is given by its color. Its intrinsic color, $(V-I)_{0} = 0.66$,
is bluer
than normal CephIIs of this period (W Vir stars) and 
similar to peculiar W Vir stars.
This classification is supported by its light curve.
In addition, Hipparcos astrometry suggests the existence of peculiar W Vir
stars are binaries (see \cite{Feast-2008}).

\section{RV Tau variables}

\begin{figure}
  \includegraphics[width=.45\textwidth]{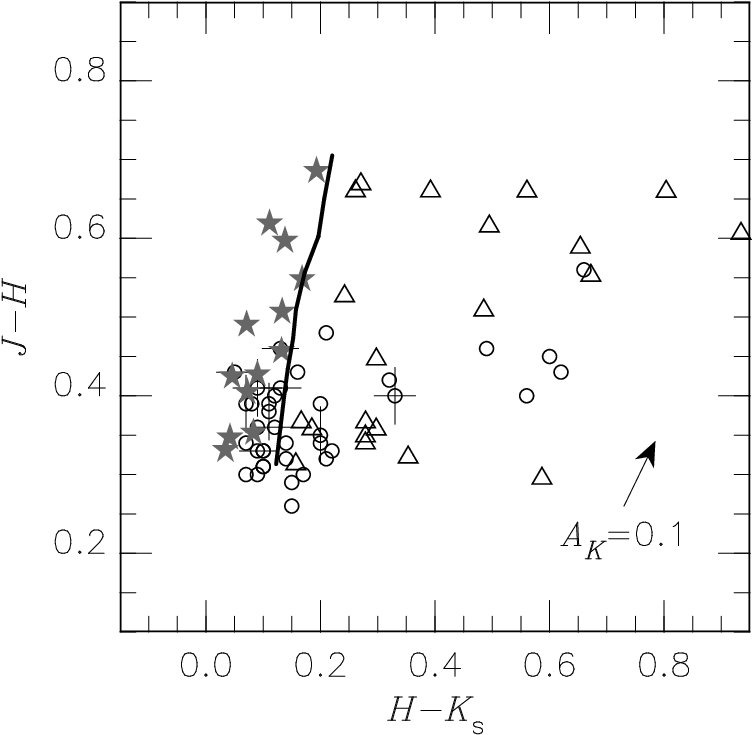}
  \caption{
Color-color diagram for RV Tau stars.
Those in the LMC are indicated by open circles,
while those in globular clusters are by star symbols.
The triangles indicates Galactic field RV stars
from \cite{LloydEvans-1985}.
Error bars are drawn for the LMC objects
only if an uncertainty significantly exceeds the size of the symbols.
The thick curve is the loci of
local giants \cite{Bessell-1988}.
\label{fig:CCD}}
\end{figure}

\begin{figure}
  \includegraphics[width=.45\textwidth]{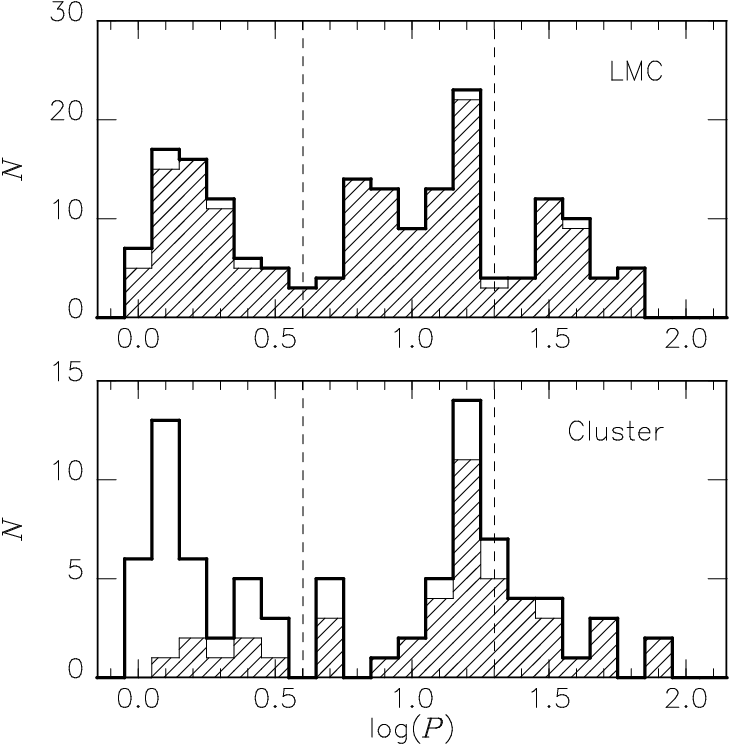}
  \caption{
Histograms of periods for the CephIIs: the top panel
is for the LMC objects from \cite{Soszynski-2008}
and the lower panel is for those in globular clusters
from a combined catalogue of \cite{Matsunaga-2006} and 
\cite{Pritzl-2003}.
The hatched areas indicate the period distribution of the objects
whose near-infrared magnitudes are discussed in this study.
Vertical lines indicate period
the divisions adopted by \cite{Soszynski-2008}.
\label{fig:Phist}}
\end{figure}

RV Tau stars in the LMC ($1.3<\log P<1.8$) do not lie on a linear extension
of the PLRs defined by shorter-period CephIIs (Fig.~\ref{fig:PLR}).
CephIIs in globular clusters in this period range, in contrast,
are co-linear with the shorter-period stars in $JHK_{\mathrm{s}}$ PL diagrams.
In the color-color diagram (Fig. \ref{fig:CCD}), the LMC stars with
significant $K_{\mathrm{s}}$ excesses lie at $(H-K_{\mathrm{s}})\sim 0.6$~mag.
Most of the others lie relatively close to the intrinsic line
taken from \cite{Bessell-1988}.
Galactic RV Tau stars from \cite{LloydEvans-1985}, triangles in
Fig.~\ref{fig:CCD}, show a rather dispersed distribution; their $J-H$ colors
range from 0.1 to 0.7 mag. While there may be some overlap between the
three populations, in general they form distinctly different groupings.
In Fig.~\ref{fig:Phist}, we plot period histograms for CephIIs in the LMC
(top) and globular clusters (bottom). For $P>20$~days, the cluster variables
seem simply the long-period tail of the distribution of W Vir stars whereas
in the LMC there is a distinct population at these periods.
RV Tau stars are
apparently a heterogeneous group and
further work is required to investigate this.

\section{Comparison with the PLR of classical Cepheids}

\begin{figure}
  \includegraphics[width=.45\textwidth]{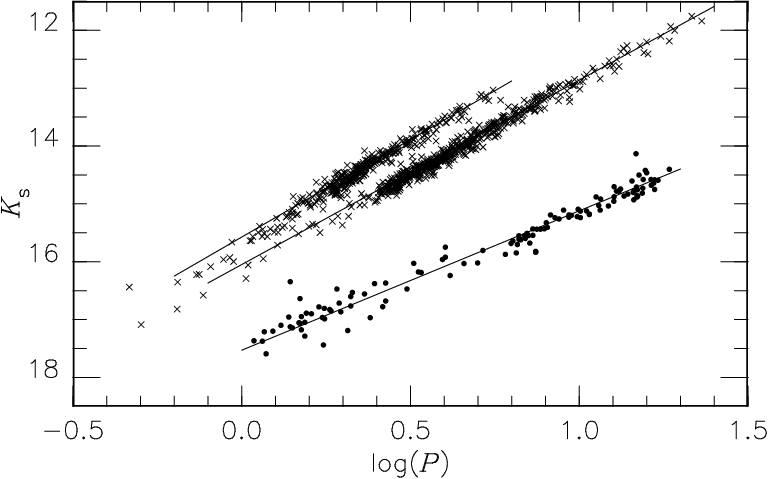}
  \caption{
PLRs of CephIIs (points) and classical Cepheids (crosses, data taken from \cite{Ita-2004}.
\label{fig:compPLR}}
\end{figure}

Since the OGLE-III project has discovered a sizable number of
CephIIs and classical Cepheids, their PLRs can be compared
directly based on a homogeneous dataset (the OGLE-III optical data and
the IRSF near-IR data). We plot the $K_{s}$ 
PLRs of both classical Cepheids and CephIIs
in Fig.~\ref{fig:compPLR}. 
The data for the classical Cepheids come from \cite{Ita-2004}.
It has long been known that the CephIIs are fainter than classical
Cepheids of the same period.
It has long been known that the CephIIs are fainter than classical Cepheids
of the same period.
Fig.~\ref{fig:compPLR} also illustrates their different PLR slopes.
These slopes are related to the mass distributions of both groups
of variables.
For the classical Cepheids, mass increases with period
while the masses of the CephIIs are almost constant with period
\cite{Matsunaga-2006}. It is also interesting that there is
a clear sequence of overtone pulsators for classical Cepheids
which has not been seen for CephIIs.

\section{Summary}

We obtained PLRs in phase-corrected $JHK_{\mathrm{s}}$ magnitudes for
a combined set
of BL Her and W Vir stars in the LMC. They have slopes consistent with
those found previously in globular clusters. 
The zero-point of the PLR is also determined based on
nearby CephIIs with pulsation parallaxes.
These has led to the 
establishment of the PLR of CephIIs as 
a distance indicator, which will be a useful tool for
investigating old stellar systems.
Since observational information for CephIIs
is gradually accumulated in a comprehensive way,
e.g.\  \cite{Matsunaga-2006} and \cite{Soszynski-2008},
a theoretical effort to explain the recent observational results
for CephIIs is highly desired.

\begin{theacknowledgments}
NM acknowledges that a part of the travel expense to Santa Fe
was supported by Hayakawa Yukio Fund operated by
the Astronomical Society of Japan.
\end{theacknowledgments}

\bibliographystyle{aipprocl} 

\end{document}